**Title**

A Nitrogen Alternative: Use of Plasma Activated Water as Nitrogen Source in Hydroponic Solution for Radish Growth

**Authors**

Vikas Rathore[1] and Sudhir Kumar Nema[1,2]

**Affiliations**

[1]Atmospheric Plasma Division, Institute for Plasma Research (IPR), Gandhinagar, Gujarat382428, India

[2]Homi Bhabha National Institute, Training School Complex, Anushaktinagar, Mumbai,400094, India

Author to whom correspondence should be addressed:

Vikas Rathore

**Email:** vikas.rathore@ipr.res.in, vikasrathore9076@gmail.com

**Graphical Abstract**

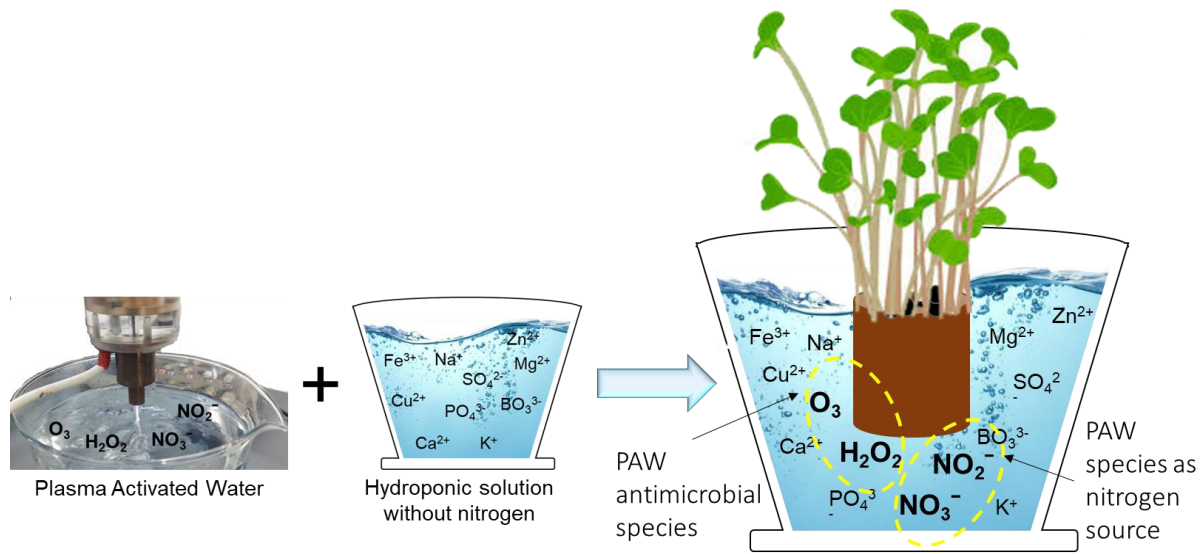

**Abstract**

The study investigates the potential of Plasma-Activated Water (PAW) as a nitrogen supplement in hydroponic cultivation (HS-N+PAW), specifically focusing on radish seed germination and plant growth. PAW, produced using a dielectric barrier discharge pencil plasma jet using air as plasma forming gas, is compared against conventional hydroponic solution (HS) and hydroponic solution without nitrogen (HS-N).

PAW treatment completely eliminates microbial growth in seeds. Radish plants cultivated with HS-N+PAW display approximately 30% and 3% longer roots compared to those grown with HS-N and HS, respectively, with shoot length increasing by ~16.5% (HS-N) and <1% (HS). Root weight sees a substantial increase of ~51% with HS-N+PAW compared to HS-N, while the increase with HS is not significant. Similarly, shoot fresh weight sees a notable increase of 50% (HS-N) and 10% (HS).

In terms of biochemical composition, radish roots show a significant increase of approximately 15.3% in soluble sugar concentration with HS-N+PAW compared to HS-N. Protein concentration in radish leaves increases by ~5.1% and ~19.0% with HS-N+PAW compared to HS-N and HS, respectively. Heightened soluble sugar and protein concentrations in HS-N+PAW-grown plants, indicating enhanced metabolic activity and nutrient uptake. However, variations in chlorophyll and carotenoid concentrations in leaves among different growth media are statistically insignificant. $H_2O_2$ concentration root and shoot remains consistent across growth media, electrolytic and phenolic leakage, along with antioxidant enzyme activities, exhibit differential responses, underscoring the impact of growth conditions on plant stress responses.

Furthermore, sensory evaluation and physical attributes analysis underscore the negative effects of nitrogen deficiency in radish plants grown with HS-N. Conversely, HS-N+PAW cultivated plants exhibit improved visual appearance, surface texture, and overall acceptance, highlighting PAW's potential as a nitrogen source for enhancing plant growth and quality in hydroponic systems.

**Keyword:** plasma activated water, nitrogen supplement, hydroponic cultivation, radish growth, biochemical analysis

# 1. Introduction

As our world undergoes rapid urbanization, traditional agriculture faces significant challenges, particularly in the context of diminishing cropland availability. With an estimated 80% of the global population projected to reside in urban areas by 2050, the pressure on agricultural systems to meet the escalating demand for food continues to mount (1, 2). The increase in global fertilizer demand will increase by 32.1%, reaching 226,150,381 Mt by 2030. This surge in demand will be driven by substantial increases in nitrogen, phosphate, and potash fertilizers, expected to rise by 37.5%, 25.8%, and 21.2%, respectively. Moreover, total fertilizer consumption is expected to be higher in developing countries like India, projected to increase by 52.8% and reach about 156,727,886 Mt by 2030 (3). In response to this challenge, alternative farming methods such as hydroponics and aeroponics have emerged as viable solutions (1, 2).

Hydroponic farming involves the cultivation of plants in nutrient-rich water solutions rather than traditional soil-based systems, offering several advantages including higher crop yields, reduced water usage, and the ability to grow crops in urban environments. Similarly, aeroponic systems suspend plant roots in the air and mist them with a nutrient solution, facilitating efficient nutrient absorption and further optimizing resource utilization (2, 4).

One innovative technology that holds promise for revolutionizing hydroponic and aeroponic agriculture is Plasma-Activated Water (PAW) (5-23). Through a process involving the exposure of water to plasma discharge, PAW is enriched with reactive oxygen and nitrogen species (RONS), imparting potent antimicrobial properties (24-30). This makes PAW an attractive alternative to chemical pesticides and fertilizers, offering sustainable solutions for plant growth enhancement and disease control (12, 31-33).

Over the years, numerous studies have investigated the efficacy of PAW in promoting plant growth, with a particular focus on radish plants (34-40). Researchers such as Guragain et al. (37), Tanakaran et al. (40), and Belov et al. (34), etc. have explored the effects of PAW treatment on radish germination and growth parameters. Their findings consistently demonstrate that PAW treatment leads to improvements in seed germination rates, seedling growth, and overall plant biomass.

Furthermore, investigations by Iwata et al. (39), Kim et al. (41), etc. have highlighted the sterilizing and microbial inactivation properties of PAW, underscoring its potential for maintaining the cleanliness of hydroponic systems. Studies by Ahn et al. (22), Gamaleev et al. (20), and Noh et al. (42) have also elucidated the synergistic effects of combining plasma technology with hydroponic systems to enhance plant growth.

The discussion above lacks comprehensive exploration of the role of nitrogen in hydroponic solutions, particularly its impact on various plant growth factors like sugar and protein concentration, chlorophyll levels, oxidative stress, and membrane integrity, as well as antioxidant levels. Additionally, there is a notable absence of detailed comparison between conventional nitrogen sources and those provided by Plasma-Activated Water (PAW). This literature gap serves as the primary motivation for our present investigation.

We aim to investigate the influence of nitrogen in hydroponic solutions on the growth and development of radish plants (*Raphanus sativus* L.). The radish plant was chosen for our present manuscript due to its reliance on the nutritional medium, whether soil-based or liquid, for its growth and yield. Various nutritional compositions significantly impact radish growth, making it an ideal candidate for studying the effects of different nutrient compositions or concentrations on plant growth. Radishes are rich in essential nutrients like carbohydrates, proteins, vitamins, and minerals (43, 44). They are widely used in salads and various cuisines globally (due to high water content ~95%), including India, where they are a staple in daily diets.

To investigate PAW as nitrogen source on radish growth, we prepared three different solutions: a conventional hydroponic solution (HS), a nitrogen-free hydroponic solution (HS-N), and a hydroponic solution utilizing PAW as the nitrogen source (HS-N+PAW). We characterized the produced PAW using a dielectric barrier discharge pencil plasma jet, examining changes in physicochemical properties and measuring the concentration of reactive oxygen nitrogen species.

Present work seeds and plant analysis including microbial inhibition in seeds, seeds germination study, plant agronomy traits, nutritional analysis, oxidative stress evaluation, electrolytic and phenolic leakage assessment, and antioxidant enzyme activity analysis. Moreover, we incorporated sensory evaluation when observing a decline in growth among the HS-N grown plants. This structured approach allows us to systematically investigate the role of nitrogen in hydroponic solutions and explore the potential benefits of utilizing PAW as a nitrogen source, thereby contributing novel insights to the field.

**2. Materials and Methods**

2.1 Plasma Activated Water (PAW) production and characterization

The production of plasma-activated water (PAW), utilized as a nitrogen source in hydroponic solutions (HS), was achieved using a pencil plasma jet (PPJ) (29, 45, 46). A detailed schematic of the PPJ setup is provided in supplementary material Figure S1. Air served as the plasma-forming gas in the PPJ setup. The characterization of the air plasma involved measuring the voltage-current of the air discharge, and the generated species/radicals were identified using optical emission spectroscopy. Further details regarding plasma diagnosis and species/radical identification can be found in our previously published reports (29, 45, 47, 48). The plasma generated in the PPJ setup exhibited filamentary dielectric barrier discharge characteristics, with a calculated plasma discharge power of approximately 10.7 W. The voltage-current waveform of the air plasma, along with the charge-voltage Lissajous figure used for plasma power calculation and emission spectra of air plasma, are shown in supplementary material Figures S2 and S3.

For the production of PAW, 500 ml of ultrapure Milli-Q water was placed in a 600 ml glass beaker. The PPJ was utilized to produce PAW by treating the ultrapure Milli-Q water with air plasma. To enhance the dissolution of air plasma reactive species/radicals in water, a quartz tube from the PPJ setup was inserted into the water to prevent the loss of reactive species/radicals before interacting with the water. The plasma-water interaction time exceeded 48 hours to achieve the desired nitrogen concentration in PAW.

Stable reactive species such as nitrate ions, nitrite ions, hydrogen peroxide, and dissolved ozone, generated in water due to plasma-water interaction, were semi-quantitatively quantified using strip tests and colorimetry detection kits (QUANTOFIX nitrite ions test strips, QUANTOFIX hydrogen peroxide test strips, VISOCOLOR alpha nitrate ions colorimetry kit, and HANNA dissolved O3 colorimetry test kit), and quantitatively using UV-visible spectroscopy. The concentration of $NO_3^-$ ions in PAW was measured using the UV screening method, where the absorbance of the sample was recorded at 220 nm, and the unknown concentration was determined using a standard $NO_3^-$ ions curve. The concentration of $NO_2^-$ ions in PAW was determined using a freshly prepared Griess reagent in an acidic region, where $NO_2^-$ ions were diazotized with a reagent made using sulfanilamide and N-(1-naphthyl) ethylenediamine dihydrochloride to produce a pink azo dye that shows maximum absorption at 540 nm. The concentration of $H_2O_2$ in PAW was determined using the titanium sulfate method, where $H_2O_2$ reacts with titanium ions in an acidic region to give a yellow color that shows maximum absorption at 407 nm. The concentration of dissolved $O_3$ in PAW was determined using the indigo colorimetry method (26, 27, 47, 49, 50). Further details regarding the measurement of these reactive oxygen-nitrogen species concentrations in PAW are provided in supplementary material appendix A1.

2.2 Radish grown using different growth medium

Radish (*Raphanus Sativus* L.) seeds were purchased from the local farmer's market in Gandhinagar district, Gujarat, India. The seeds were thoroughly washed with sterile ultrapure Milli-Q water to remove any settled dirt and dried using sterile paper towels. Plasma-activated water (PAW) was utilized as a disinfection solution to inhibit microbial growth on the seed surface, and its efficacy was compared with hydroponic solutions without nitrogen (HS-N) and with nitrogen (HS). The measurement of colony-forming units (CFU) was employed to determine the microbial count on the seed surface. The microbial reduction on the seed surface was calculated using the following expression:

$$\% \, microbial \, reduction = \left| \frac{(CFU/g)_{sample} - (CFU/g)_{control}}{(CFU/g)_{control}} \right| \times 100 \qquad (1)$$

Where "sample" represents PAW, HS-N, HS, and "control" denotes ultrapure Milli-Q water.

The experimental schematic shown the use of PAW as a nitrogen source in hydroponic solution (HS-N+PAW) for the germination and growth of radish, in comparison to hydroponic solution (HS) and hydroponic solution without nitrogen, is shown in Figure 1. All growth solutions were prepared using ultrapure Milli-Q water and laboratory-grade chemicals. The HS growth medium was composed of micromaterials such as calcium nitrate ($CaNO_3 \cdot 4H_2O$), potassium nitrate ($KNO_3$), magnesium sulfate ($MgSO_4 \cdot 7H_2O$), ammonium dihydrogen phosphate ($NH_4H_2PO_4$), and micronutrient disodium iron ethylene dinitrilo tetraacetate ($Na_2Fe.EDTA$), boric acid ($H_3BO_3$), manganese sulfate ($MnSO_4 \cdot 4H_2O$), zinc sulfate ($ZnSO_4 \cdot 7H_2O$), copper sulfate ($CuSO_4 \cdot 7H_2O$), and ammonium heptamolybdate $((NH_4)_6Mo_7O_{24})$ as reported by Li et al. (51). Nitrogen-containing salts such as calcium nitrate $(CaNO_3 \cdot 4H_2O)$, potassium nitrate $(KNO_3)$, etc., were not added to the HS-N medium. PAW was added to the HS-N medium as a nitrogen source to prepare the HS-N+PAW medium. The concentration of calcium, potassium, and phosphorus in the growth media (HS-N and HS-N+PAW) was maintained using potassium chloride (KOH), calcium chloride $(CaCl_2)$, and potassium dihydrogen phosphate $(KH_2PO_4)$. The pH of the solution was maintained around 6.5 for all growth media using sulfuric acid $(H_2SO_4)$ and potassium hydroxide (KOH). The detailed chemical composition of the growth media is provided in Table S1 of Appendix A2 of the supplementary material.

Figure 1 shown the production of PAW and the preparation of different growth media (HS-N, HS, and HS-N+PAW), along with images of radish seed germination (Day 2) and the growth of radish plants at different time intervals (Day 7 to Day 25).

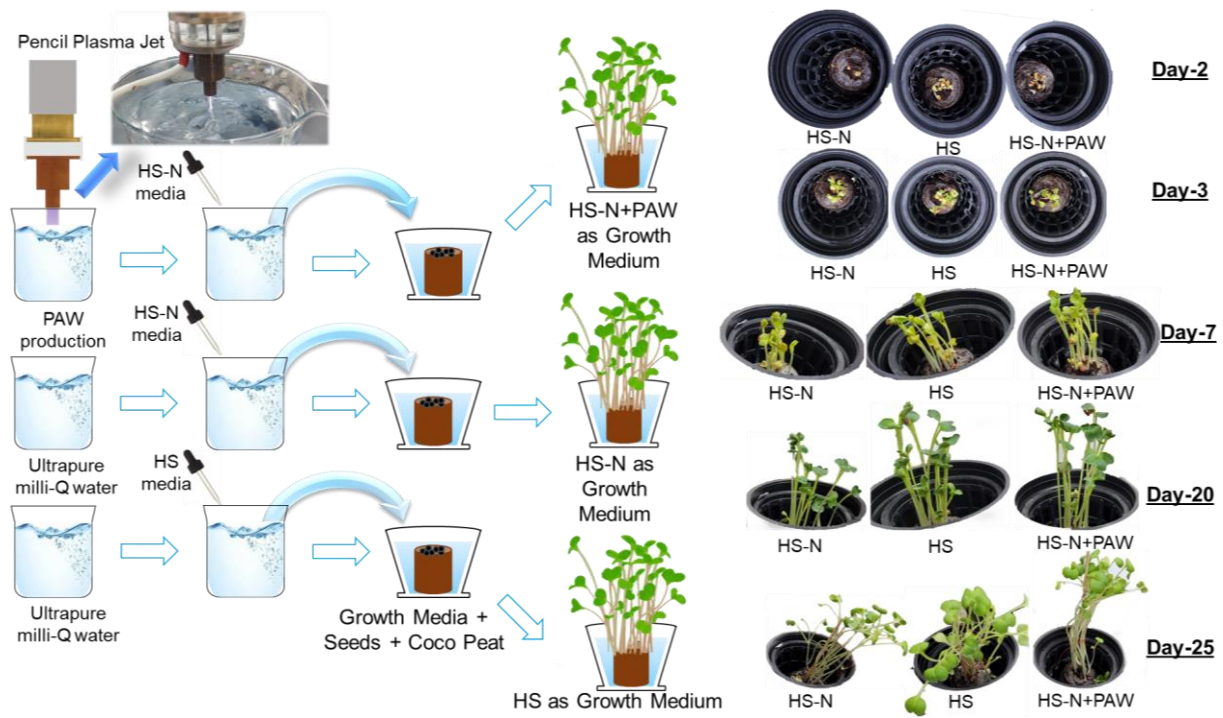

**Figure 1**. Schematic of the use of plasma-activated water (PAW) as a nitrogen source for hydroponic solution (HS) as the growth medium for radish plants. HS-N: hydroponic solution without nitrogen, HS-N+PAW: hydroponic solution with PAW as the nitrogen source.

2.3 Germination and plant growth analysis

To examine the influence of different growth media on radish seed germination and plant growth, a total of 360 radish seeds (3 groups × 3 replicates × 40 seeds) were utilized and evenly distributed among three groups: HS-N, HS, and HS-N+PAW. Each group was further subdivided into three sets of 40 seeds to ensure result replication. Following seed disinfection as described earlier, seeds were placed in growth media-absorbed coco peat pellets, as shown in Figure 1, and germination and plant growth were monitored at regular intervals.

Seed germination was observed up to day 5 under dark conditions at a temperature of 24°C and 40% relative humidity, while plant growth was assessed up to day 30 under a light-dark cycle of 16-8 hours, with an irradiance level of 44 W m$^{-2}$, temperature at 24°C, and relative humidity at 40%.

To analyze the physical attributes and biochemical composition of radish plants, three-week-old plants were harvested. Physical attributes such as root and shoot length, fresh weight, and biochemical analyses including soluble sugar and protein content, chlorophyll concentration in leaves, oxidant levels such as hydrogen peroxide, membrane integrity, and antioxidant activity of leaves and roots were studied. During the fourth week of the experimental duration, radish plants grown using HS-N exhibited signs of growth decline, wilting, dehydration, and loss of vitality. Therefore, sensory characteristics such as visual appearance and surface texture, along with physical attributes of HS-N grown plants, were compared to those of HS and HS-N+PAW grown plants.

The chlorophyll ($C_a$ and $C_b$) and carotenoid ($C_{c+x}$) concentrations in fresh radish leaves were measured using UV-visible spectroscopy (SHIMADZU UV-2600) (52). 0.1 g of fresh radish leaves were ground with an ice-cooled mortar and pestle and added to a freshly prepared aqueous acetone solution (80% volume/volume). The mixture was centrifuged, and the absorbance of the supernatant was recorded at 470 nm, 647 nm, and 663 nm to measure chlorophyll and carotenoid concentrations using the following expressions (see Appendix A3 of the supplementary material for detailed procedures) (5, 52):

$$C_a \left(\frac{\mu g}{ml}\right) = 12.25 A_{663} - 2.79 A_{647} \quad (2)$$

$$C_b \left(\frac{\mu g}{ml}\right) = 21.5 A_{647} - 5.1 A_{663} \quad (3)$$

$$C_{c+x} \left(\frac{\mu g}{ml}\right) = (1000 A_{470} - 1.82 C_a - 85.02 C_b) / 198 \quad (4)$$

The concentration of soluble sugars and proteins in leaves and roots was determined following the procedures outlined by Pons et al. (53) and Lowry et al. (54), respectively. For sugar estimation, 0.1 g of fresh leaves or roots was added to a warm aqueous ethanol solution (80% volume/volume) and heated in a boiling water bath for sugar extraction. After cooling, a 0.2% freshly prepared ice-cold anthrone reagent was added to the extract supernatant, and absorbance was recorded at 590 nm. Sugar quantification was performed using a standard curve prepared with different concentrations of dextrose.

Soluble proteins were extracted from radish plant leaves or roots using lysis buffer (protein extraction buffer). Ice-cooled 0.1 g of ground plant material (leaves or roots) was mixed with lysis buffer, followed by the addition of 2N sodium hydroxide to the extracted protein solution. The solution was hydrolyzed in a boiling water bath, cooled, and then combined with freshly prepared complex reagent (mixture) and Folin reagent. The developed color absorbance was recorded at 750 nm. A bovine serum albumin standard curve was utilized to quantify protein concentration (see Appendix A3 of the supplementary material for detailed sugar and protein estimation procedures).

Oxidative stress in the form of $H_2O_2$ was determined using the potassium iodide method (55). $H_2O_2$ was extracted from 0.1 g of leaves or roots using 0.1% trichloroacetic acid. In the $H_2O_2$ extract supernatant, 100 mM potassium phosphate buffer and 1 M potassium iodide reagent were added, and absorbance was recorded at 390 nm after an hour of incubation. A standard $H_2O_2$ curve was employed for $H_2O_2$ quantification in leaves or roots (see Appendix A3 of the supplementary material for detailed $H_2O_2$ estimation procedures).

Electrolytic and phenolic leakage from fresh radish leaves were measured by electrical conductivity and absorbance (260 nm) of 0.1 g of half-cut leaves in ultrapure Milli-Q water (56) (see Appendix A3 of the supplementary material for more information).

Antioxidant enzymes were extracted from 0.1 g of leaves and roots using ice-cooled extraction buffer containing 0.8 M phosphate buffer, 0.2 N EDTA solution, ascorbic acid, and ultrapure Milli-Q water (57). The activities of SOD (superoxide dismutase) and CAT (catalase) in leaves and roots were determined using the procedures provided in the SOD Determination Kit (Sigma-Aldrich) and Catalase Assay Kit (Cayman Chemical), respectively. APX activity was determined by adding a reaction mixture (50 mM potassium phosphate, 0.5 mM ascorbate, 0.1 mM $H_2O_2$, and 0.1 mM EDTA) to the enzyme extract and measuring the decrease in absorption from 10 to 30 s at 290 nm (attenuation coefficient 2.8 $mM^{-1}$ $cm^{-1}$) (58). POD activity was determined by adding a reaction mixture containing phosphate buffer, guaiacol solution, and $H_2O_2$ to the enzyme. The reaction was initiated upon adding the enzyme to the reaction mixture, and absorbance was measured at 436 nm (59). POD activity was calculated using the following expression:

$$POD\ activity\ (U\ l^{-1}) = \frac{500}{\Delta t} \tag{5}$$

Where Δt is the time required to increase absorbance by 0.1. (see Appendix A3 of the supplementary material for detailed procedures).

2.4 Data analysis

All experiments were replicated at least three times, and results were expressed as mean ± standard deviation in tables and graphs/plots. Statistical analysis of the results was performed using analysis of variance followed by post-hoc test (Fisher's least significant difference (LSD)) with a statistical significance level of 95% (p-value 0.05).

**3. Results**

3.1 PAW properties

Table 1 presents the properties of Plasma-Activated Water (PAW) used as a nitrogen supplement in hydroponic solutions. The prepared PAW exhibited significant acidity, with a pH decrease from 6.9 (in ultrapure milli-Q water, the control) to 1.9. This drop in pH indicates the formation of acidic compounds during plasma-liquid interaction, such as nitrous and nitric acid, reflected in the form of nitrates and nitrites concentration in Table 1. Additionally, PAW displayed a consistently high oxidizing potential (Oxidation-Reduction Potential - ORP) close to 700 mV, indicating that the dissolved species are primarily oxidizing in nature. The reactive oxygen-nitrogen species (RONS) listed in Table 1 are also oxidizing, with standard electrode potentials (E°) ranging from 0.9 V (for $NO_3^-$ ions) to 2.1 V (for Dissolved $O_3$). The increasing electrical conductivity (EC) and total dissolved solids (TDS) suggest the presence of dissolved inorganic ions in PAW, such as $NO_3^-$ ions, $NO_2^-$ ions, and $H^+$ ions. Although dissolved $O_3$ and $H_2O_2$ do not directly contribute to TDS and EC, their reaction with other dissolved species generates reactive ions that significantly enhance the TDS and EC of PAW. Similar decreases in pH and/or increases in oxidation-reduction potential and electrical conductivity following plasma-water interaction have been reported in previous studies by Lamichhane et al. (32),

Wang et al. (60), TAKANO et al. (9), Date et al. (11), Takahashi et al. (13), Tephiruk et al. (61), and Gott et al. (62), etc.

The concentration of $NO_3^-$ ions (> 813 mg L-1) in PAW is notably high compared to other RONS ($NO_2^-$ ions, $H_2O_2$, and dissolved $O_3$), as indicated in Table 1. This is attributed to the high stability of $NO_3^-$ ions in PAW, whereas other RONS react with each other to form stable $NO_3^-$ ions. For instance, $NO_2^-$ ions react with $H_2O_2$ and dissolved $O_3$ to produce $NO_3^-$ ions (see supplementary material Appendix A4 for reaction mechanism). The formation of $H_2O_2$ and $NO_3^-$ ions after plasma treatment has also been reported by Maniruzzaman et al. (63), Lamichhane et al. (32), Wang et al. (60), TAKANO et al. (9), Date et al. (11), Song et al. (12), Takahashi et al. (10), Tephiruk et al. (61), Ruamrungsri et al. (15), and Gott et al. (62), etc.

Table 1. Physicochemical properties and reactive species concentration in PAW. Different symbols in groups ($\mu \pm \sigma$) indicate statistically significant differences ($p < 0.05$) among them.

| Sample | pH | ORP (mV) | TDS (ppm) | EC ($\mu S\ cm^{-1}$) | $NO_2^-$ (mg L$^{-1}$) | $NO_3^-$ (mg L$^{-1}$) | $H_2O_2$ (mg L$^{-1}$) | $O_3$ (mg L$^{-1}$) |
|---|---|---|---|---|---|---|---|---|
| Control | $6.9^* \pm 0.1$ | $260^* \pm 10$ | $0^* \pm 0$ | $1^* \pm 0$ | $0.0^* \pm 0.0$ | $0.0^* \pm 0.0$ | $0.0^* \pm 0.0$ | $0.0^* \pm 0.0$ |
| PAW | $1.9^o \pm 0.2$ | $690^o \pm 45$ | $4305^o \pm 213$ | $10112^o \pm 1251$ | $25.9^o \pm 3.5$ | $813.3^o \pm 41.2$ | $9.5^o \pm 1.7$ | $1.7^o \pm 0.3$ |

3.2 Microbial analysis and seeds germination

Figure 2(a) shown the reduction in microbial load on radish seeds after exposure to HS, HS-N, and PAW compared to the control. PAW-treated seeds exhibited complete microbial load reduction (100%) compared to the control (ultrapure milli-Q water). This is attributed to the high concentration of reactive oxygen species like H2O2 and dissolved O3, known for their efficacy in microbial inactivation. Furthermore, the acidic environment generated by formed nitrous and nitric acid also supports the antimicrobial activity of PAW. The antimicrobial activity of plasma treatment of liquid for disinfection to enhance germination and/or plant growth has been previously reported by Belov et al. (34), Iwata et al. (39), Saito et al. (8), Abbaszadeh et al. (6), Okumura et al. (21), Date et al. (11), Carmassi et al. (16), Park et al. (19), Gamaleev et al. (20), and Takaki et al. (23), etc.

PAW treatment also aids in removing naturally occurring wax from the seed surface due to the interaction of high oxidizing species with the wax, resulting in the hydrophobic surface of the seeds becoming hydrophilic (5). Consequently, the rate of water absorption by seeds increases, leading to rapid germination. This enhancement in germination was also observed in PAW-treated seeds (Figure 2(b)). On day 2 of germination, the seeds grown using HS-N+PAW as a growth medium exhibited approximately 14% and 10% higher germination rates compared to HS-N and HS growth media, respectively. As shown in Figure 2(b), this increase was not statistically significant. However, the growth medium did not play a role in germination; this increase in germination percentage was due to PAW treatment. Jeon et al. (7) and Song et al. (17) have also reported enhancements in the germination rate of hydroponic systems after plasma discharge in liquid. The positive combined impact of plasma and PAW on radish seed germination has been previously reported by Sivachandiran et al. (36).

3.3 Physical growth attributes

To comprehend the overall development and productivity of radish plants when Plasma-Activated Water (PAW) was used as a nitrogen source in the growth medium, the length and fresh weight of the radish plants were analyzed.

The length and fresh weight of radish plants grown using PAW as a nitrogen source in hydroponic media (HS-N+PAW), either higher or equivalent compared to conventional hydroponic media with nitrogen (HS) and without nitrogen (HS-N), are shown in Figure 2 (c, d). The root length of plants grown using HS-N+PAW was approximately 30% and 3% higher than those grown using HS-N and HS, respectively. In the case of shoot length, this increase was about 16.5% (HS-N) and less than 1% (HS), respectively. A study by Ruamrungsri et al. (15) supports the results of this investigation, as they found that green oak lettuces exhibited higher growth in plasma nitrate solution compared to non-nitrate solutions. Moreover, when compared to conventional nitrate solutions, this growth was non-significant, suggesting the use of PAW as a nitrogen alternative for the growth of green oak lettuces.

In the fresh weight analysis, it was observed that the root weight of radish plants exhibited a significant increase of approximately 51% and a non-significant increase of about 7% when cultivated using the HS-N+PAW medium compared to those grown under HS-N and HS media, respectively. Furthermore, there was a notable enhancement in the fresh weight of the shoot by 50% and 10% when subjected to HS-N and HS growth media, respectively. However, this increase was statistically insignificant when comparing them. Takano et al. (9) also emphasize the importance of nitrogen production in the form of acid acting as a fertilizer. They produced underwater bubble discharge that enhanced the height and dry weight of Brassica rapa var. perviridis in hydroponic cultivation compared to the control; however, these increases were statistically insignificant.

The improvements in root length, shoot length, fresh weight of the root, and fresh weight of the shoot all demonstrate significant improvements, indicating the effectiveness of HS-N+PAW as a growth medium in promoting robust plant growth and biomass accumulation.

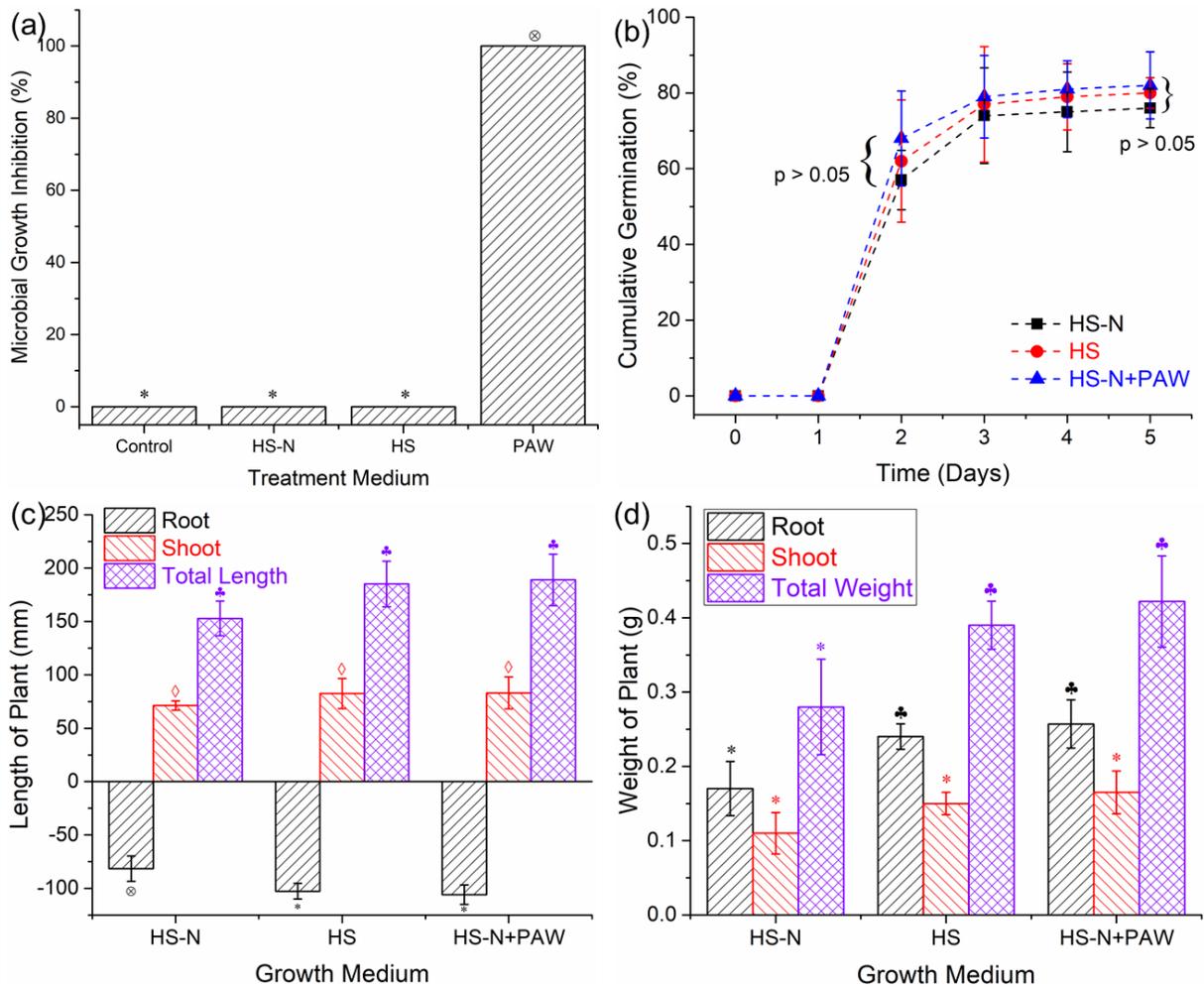

**Figure 2.** (a) Microbial growth on radish seeds after treatment with HS-N and HS media and PAW, (b) Cumulative germination of radish seeds using HS-N, HS, and HS-N+PAW as growth medium, (c, d) Length and fresh weight of radish plants growing using HS-N, HS, and HS-N+PAW as growth medium at the end of 3-weeks. Different symbols on the graph represent statistically significant differences ($p < 0.05$) between the groups ($\mu \pm \sigma$).

3.4 Nutritional analysis

By analyzing the protein and sugar content in roots and leaves, along with leaf chlorophyll concentration, significant insights into plant growth dynamics can be obtained. The variation in soluble sugar, soluble protein, and chlorophyll concentration when using different growth media is illustrated in Figure 3. The soluble sugar concentration in radish roots exhibited a notable increase of approximately 15.3% when grown using the HS-N+PAW medium compared to the HS-N medium (Figure 3(a)). Conversely, the sugar concentration in leaves showed an insignificant increase of about 4.1% compared to the HS-N medium, whereas a decline of 3.8% in sugar concentration was observed compared to those grown in the HS medium. However, this increase or decline in sugar concentration was not statistically significant compared to each other. A higher concentration in HS-N+PAW grown radish plants signifies better metabolic activity and carbohydrate storage, especially for energy production and various physiological processes.

The protein concentration of radish leaves grown using the HS-N+PAW medium showed an increase of approximately 5.1% and 19.0% compared to HS-N and HS growth media, respectively (Figure 3(b)). However, in the case of radish roots, there was a decline of about 5.4% and 9.1% compared to plants grown using HS-N and HS media, respectively. Elevated protein concentration in plant leaves indicates better plant health and efficacy in utilizing nutrient assimilation. Moreover, it signifies a better ability to utilize available resources for plant growth and development.

The variation in chlorophyll ('$C_a$' and '$C_b$') and carotenoid ('$C_{c+x}$') concentrations in leaves of radish grown using different growth media (HS-N, HS, and HS-N+PAW) was statistically insignificant ($p > 0.05$), as shown in Figure 3(c). This suggests that the nutrient composition provided by each growth medium did not significantly impact the photosynthetic pigments in the leaves. The consistent chlorophyll and carotenoid concentrations imply that the physiological processes related to photosynthesis were not markedly influenced by the choice of growth medium. The chlorophyll ratio ($Ca\ Cb^{-1}$) suggests the plant's adaptation to different light conditions and its overall photosynthetic efficiency. No significant variation in the chlorophyll ratio suggests that the plants maintain a consistent balance between these two chlorophyll pigments regardless of the growth medium.

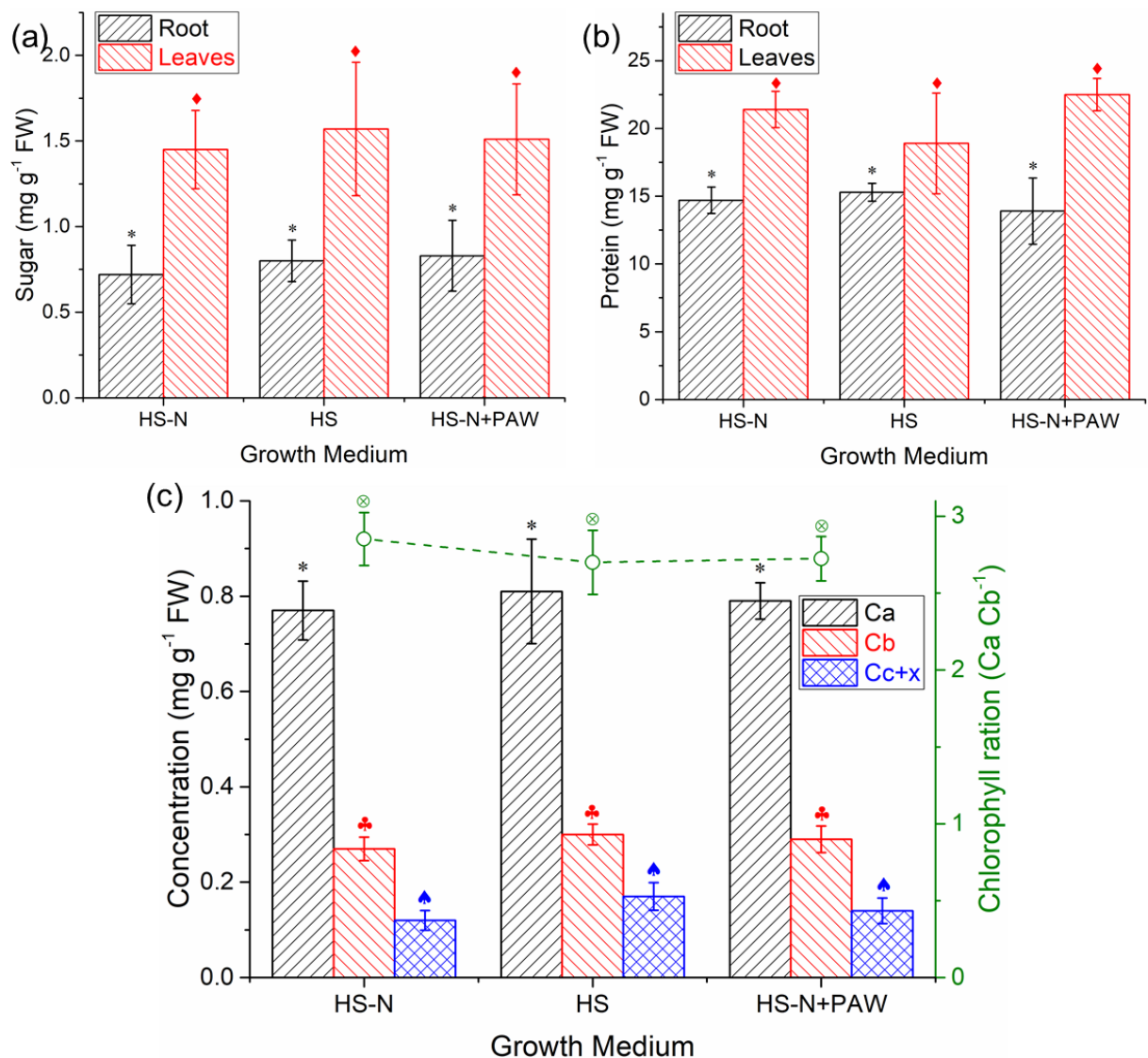

**Figure 3.** Nutritional analysis ((a) soluble sugar, (b) soluble protein, and (c) chlorophyll concentration) of radish plants grown using HS-N, HS, and HS-N+PAW as growth medium. Different symbols on the graph represent statistically significant differences ($p < 0.05$) between the groups ($\mu \pm \sigma$).

3.5 Oxidative stress, leakages, and antioxidant analysis

The oxidative stress ($H_2O_2$ concentration) in roots and leaves grown using different growth media did not show any statistically significant variation ($p > 0.05$), as shown in Figure 4(a).

However, the observed H2O2 concentration in roots and leaves of radish plants grown in HS medium was relatively low compared to other growth media. This suggests that oxidative stress in plants is not dependent on the nutrient composition or environmental conditions provided by the growth media. It highlights the plant's ability to maintain cellular homeostasis under varying growth conditions. Radish plants grown in HS medium exhibited lower H2O2 levels, hinting at its potential to alleviate oxidative stress. This suggests that HS medium contains components fostering a healthier environment for radish growth, thereby bolstering their vitality.

The evaluation of electrolytic (EC: electrical conductivity) and phenolic leakage serves as a reliable indicator of membrane integrity in plant leaves. Membrane integrity is crucial for maintaining cellular functions and protecting against stress-induced damage. The use of HS-N+PAW as a growth medium results in lower electrolytic and phenolic leakage compared to HS and HS-N media, indicating potential benefits for plant health, as shown in Figure 4(b). While the differences in electrolytic leakage between growth media are not statistically significant, except for phenolic leakage for the HS-N medium, which was significantly higher compared to the HS-N+PAW medium.

Figure 4 (c) showed the antioxidant activity of enzymes present in radish plant roots and shoots grown using different growth media. These antioxidant enzymes, like Superoxide Dismutase (SOD), Peroxidase (POD), Catalase (CAT), and Ascorbate Peroxidase (APX), in radish plants are crucial for maintaining cellular health and resilience against oxidative stress. SOD and CAT work to neutralize harmful reactive oxygen species (ROS), while POD and APX detoxify hydrogen peroxide, preventing cellular damage. Their combined efforts ensure optimal growth, development, and stress tolerance in radish plants, highlighting the significance of antioxidant defense mechanisms in plant physiology (64-66).

The consistent level of SOD enzyme in radish plant leaves with different growth media signifies its robustness and stability in leaf tissues, protecting leaf cells from oxidative damage and ensuring their proper functionality and health (64, 65). The lower SOD activity in the roots compared to those grown in HS and HS-N media could be due to the use of PAW as a nitrogen source and highlights the importance of growth conditions on SOD activity. Lower activity also signifies a lower level of harmful ROS since elevated oxidative stress results in enhanced SOD activity to prevent cellular damage.

The consistent CAT enzyme activity observed in both leaves and roots of radish plants across various growth media signifies the resilience of their antioxidant defense system. This stability suggests that the choice of growth media does not significantly influence the plant's ability to detoxify hydrogen peroxide, a crucial aspect of oxidative stress management.

The lower POD activity in leaves of radish plants cultivated in HS-N medium compared to HS and HS-N+PAW medium suggests marginally low antioxidant defense mechanisms in response to nutrient availability (67). Conversely, the notably higher POD activity in roots grown in HS medium indicates a distinct response to nutrient composition, potentially as a protective mechanism against environmental stressors relative to HS-N and HS-N+PAW medium.

The APX activity of roots of radish grown in HS-N+PAW medium was marginally higher (statistically insignificant, $p > 0.05$), and in leaves, it was significantly lower compared to HS-N and HS medium. This suggests that the roots might have been better at dealing with oxidative stress when given the HS-N+PAW medium (68). However, in the leaves, low APX activity signifies a low level of oxidative stress. These differences show how plants can respond differently to changes in the nitrogen source present in the growth medium.

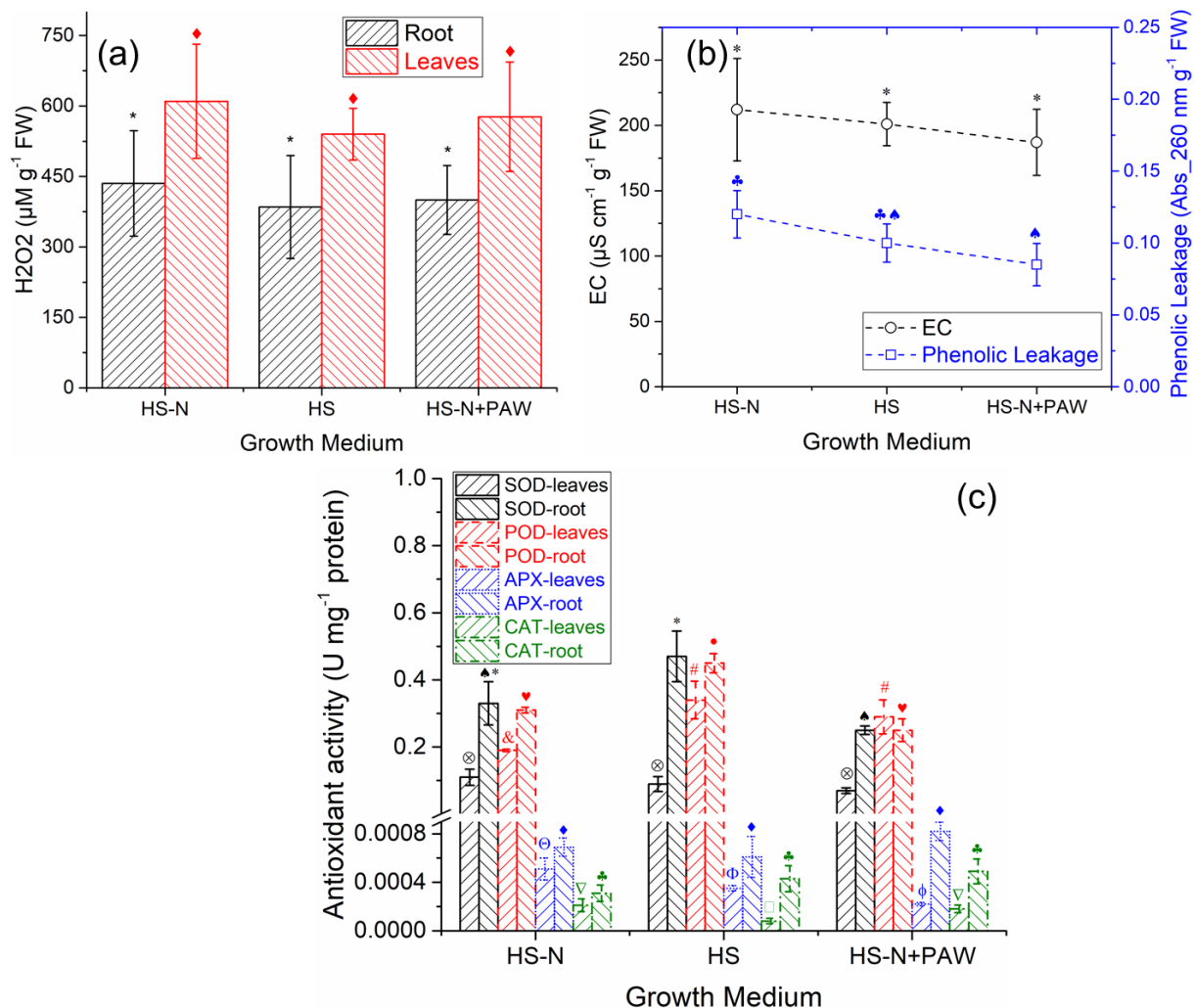

**Figure 4.** (a) H2O2 concentration in roots and leaves, (b) electrolytic and phenolic leakage from radish plant leaves, and (c) antioxidant activities (SOD, POD, APX, and CAT) of radish plant roots and leaves grown using HS-N, HS, and HS-N+PAW as growth medium. Different symbols on the graph represent statistically significant differences ($p < 0.05$) between the groups (μ ± σ).

3.6 Role of nitrogen in sensory and physical attributes

The significance of nitrogen in plant growth becomes apparent during the end phase of radish growth (Day 27 to 30). As the HS-N medium does not contain any nitrogen and was used as a growth medium for radish cultivation, initially, the plant shows steady growth since it absorbs the required nitrogen from coco peat pellets. As the plant biomass increases with time, it requires more and more nitrogen for growth, which cannot be fulfilled by the HS-N medium, resulting in radish plants showing a decline in growth, dehydration, and loss of vitality, etc. (69, 70)

This decline in growth was evaluated based on sensory and physical attributes analysis (Figures 5 and 6) and compared with cultivated radish plants in HS and HS-N+PAW medium. An image of radish plants grown using HS-N medium and HS-N+PAW medium is shown in supplementary material Figure S4. Sensory evaluation showed that radish plants cultivated in HS-N medium had very poor visual appearance, surface texture, and leaf freshness (69, 70). Hence, the use of HS-N medium was beyond acceptable limits. Similar results were observed in the length and fresh weight analysis of radish plants cultivated in HS-N medium compared to HS and HS-N+PAW medium. There was a substantial decline in the length and fresh weight (root and shoot) of plants grown in HS-N medium compared to those cultivated in media containing nitrogen. Moreover, the length and fresh weight (root + shoot) of plants cultivated in HS-N+PAW medium were marginally higher compared to the HS medium (statistically insignificant, $p > 0.05$), signifying the potential of PAW to be used as a nitrogen source in hydroponic solutions as a growth/cultivation medium.

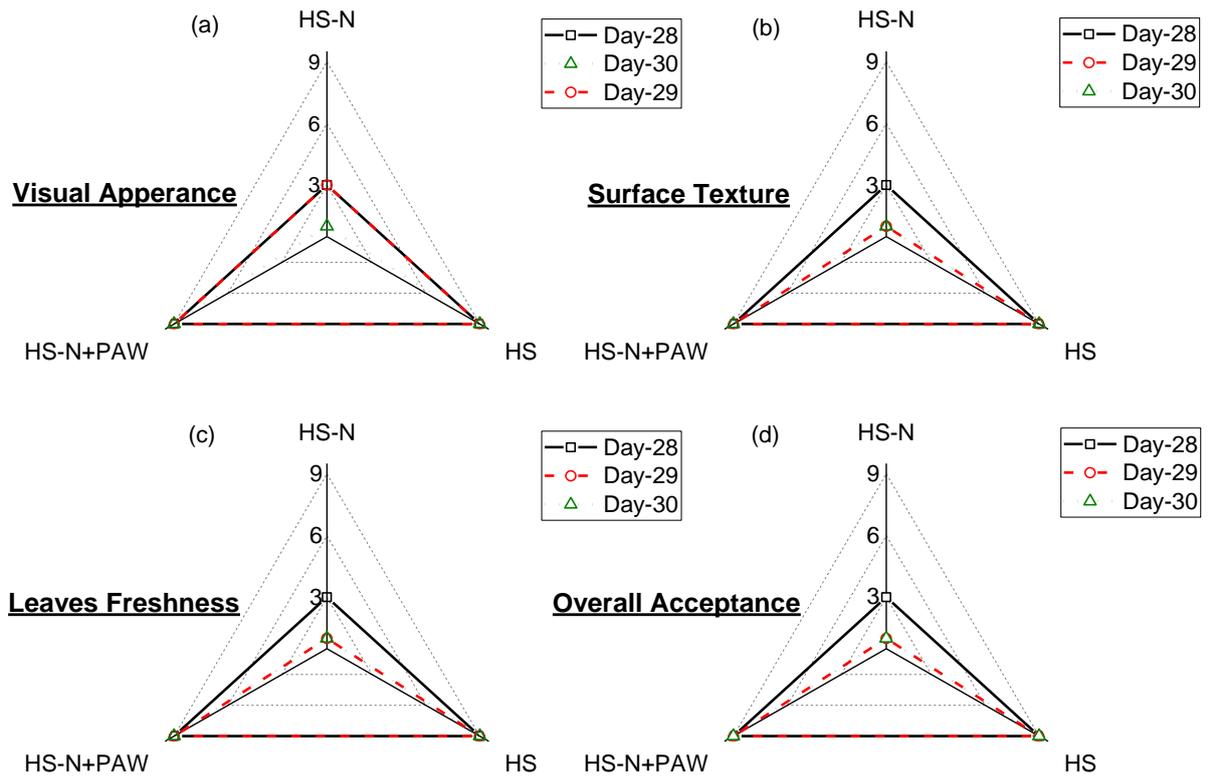

**Figure 5.** Sensory evaluation ((a) visual appearance, (b) surface texture, (c) leaf freshness, and (overall acceptance)) of radish plants grown using HS-N, HS, and HS-N+PAW as growth medium from days 28 to 30.

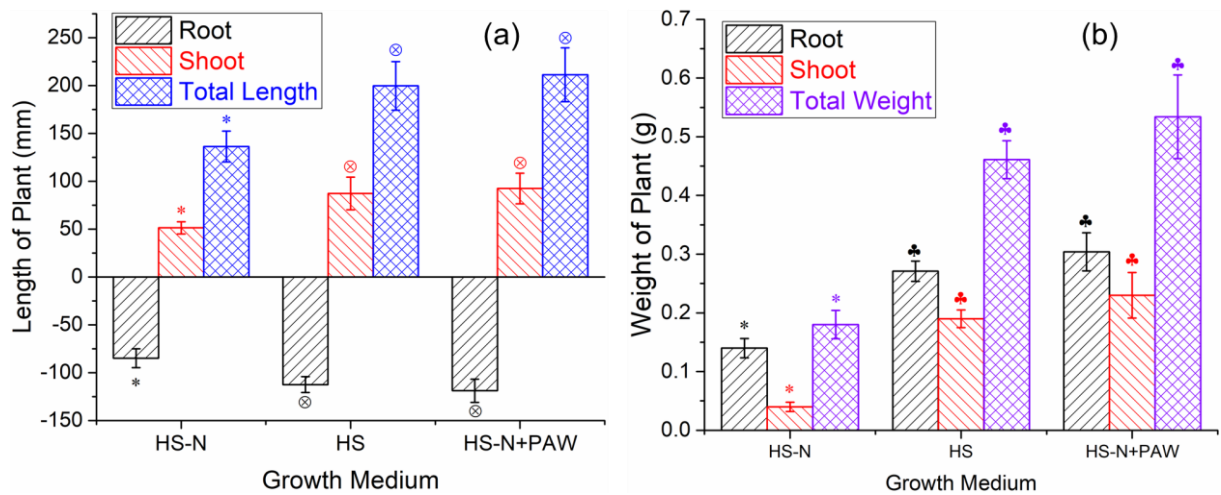

**Figure 6.** Length and fresh weight of radish plants growing using HS-N, HS, and HS-N+PAW as growth medium at the end of 30 days. Different symbols on the graph represent statistically significant differences ($p < 0.05$) between the groups ($\mu \pm \sigma$).

## 4. Discussion

The results presented in this study demonstrate the potential of Plasma-Activated Water (PAW) as a nitrogen supplement in hydroponic solutions, offering insights into its physicochemical properties and effects on microbial load reduction, seed germination, and plant growth attributes of radish plants.

PAW exhibited significantly acidic properties with a pH decrease to 1.9, indicating the formation of acidic compounds such as nitrous and nitric acid (46, 47, 50, 71). This acidic environment, coupled with high oxidizing potential, resulted in a substantial reduction in microbial load on radish seeds, enhancing seed germination rates and promoting plant growth. Additionally, PAW treatment led to the removal of naturally occurring wax from seed surfaces, thereby increasing water absorption and facilitating rapid germination (5).

When used as a nitrogen source in hydroponic media, PAW demonstrated promising effects on the physical growth attributes of radish plants. Radish plants grown with PAW exhibited comparable or enhanced root and shoot length and fresh weight compared to conventional hydroponic media with or without nitrogen supplementation. These findings suggest the potential of PAW as an alternative nitrogen source for hydroponic cultivation, offering similar or improved growth outcomes (6, 9-11, 16, 17, 61).

Furthermore, the nutritional analysis revealed significant variations in soluble sugar and protein concentrations in radish plants grown with PAW, indicating enhanced metabolic activity and nutrient assimilation. However, chlorophyll and carotenoid concentrations remained consistent across different growth media, suggesting that the photosynthetic capacity of the plants was not significantly influenced by the choice of growth medium (5, 64, 65).

Evaluation of oxidative stress indicators and antioxidant activities revealed insights into the plant's response to different growth media. While $H_2O_2$ concentration did not vary significantly among growth media, electrolytic and phenolic leakage, as well as antioxidant enzyme activities, showed differential responses, highlighting the importance of growth conditions in modulating plant stress responses (64, 65).

Finally, sensory evaluation and physical attributes analysis demonstrated the detrimental effects of nitrogen deficiency in radish plants grown with nitrogen-free hydroponic medium (HS-N), leading to reduced growth and vitality compared to plants grown with nitrogen supplementation (69, 70). PAW-treated plants showed improved visual appearance, surface texture, and overall acceptance, further highlighting the potential of PAW as a nitrogen source for enhancing plant growth and quality in hydroponic systems.

## 5. Conclusion

This study demonstrated the potential of Plasma-Activated Water (PAW) as a viable nitrogen supplement in hydroponic cultivation. Through comprehensive analysis, PAW demonstrated significant efficacy in reducing microbial load, promoting seed germination, and enhancing physical growth attributes in radish plants when used as a nitrogen source in hydroponic solution (HS-N+PAW). Additionally, radish plants grown with HS-N+PAW showed improvements in soluble sugar and protein concentrations, indicating enhanced metabolic activity. While chlorophyll and carotenoid concentrations remained stable, suggesting minimal impact on photosynthetic capacity, variations in oxidative stress indicators and antioxidant activities highlighted the influence of growth conditions on plant responses. Overall, these findings suggest that PAW holds promise as a sustainable and effective nitrogen source in hydroponic agriculture, offering potential benefits for improving crop yield and quality. Moreover, further research is necessary to optimize its application and explore its long-term effects on plant growth and development.

**Conflict of interests**

The authors declare that there are no conflicts of interests.